\documentclass[sigconf]{acmart}

\usepackage{enumitem} 
\usepackage{algorithm}
\usepackage[noend]{algpseudocode} 
\usepackage{multirow} 
\usepackage{listings}
\usepackage{url} 

\usepackage{bm}

\newcommand{\algrule}[1][.2pt]{\par\vskip.5\baselineskip\hrule height #1\par\vskip.5\baselineskip}

\AtBeginDocument{%
  \providecommand\BibTeX{{%
    \normalfont B\kern-0.5em{\scshape i\kern-0.25em b}\kern-0.8em\TeX}}}

\setcopyright{acmcopyright}
\copyrightyear{20201}
\acmYear{2021} 

\acmConference[WWW'21]{The Web Conference 2021}{April 19--23, 2021}{Ljubljana, Slovenia}
\acmBooktitle{Proceedings of the 2021 World Wide Web Conference (WWW ’21),, April 19--23, 2021, Ljubljana, Slovenia}
\acmPrice{}
\acmDOI{}
\acmISBN{}

\begin{document}

\title{Persona2vec: A Flexible Multi-role Representations Learning Framework for Graphs}



\author{Jisung Yoon}
\affiliation{%
  \institution{Department of Industrial Management Engineering, Pohang University of Science and Technology}
  \city{Pohang}
  \postcode{37673}
  \country{Korea}
}

\author{Kai-Cheng Yang}
\affiliation{%
  \institution{Center for Complex Networks and Systems Research; Luddy School of Informatics, Computing, and Engineering, Indiana University}
  \city{Bloomington}
  \state{IN}
  \country{USA}}

\author{Woo-Sung Jung}
\affiliation{%
  \institution{Department of Physics and Department of Industrial Management Engineering, Pohang University of Science and Technology}
  \city{Pohang}
  \postcode{37673}
  \country{Korea}
}

\author{Yong-Yeol Ahn}
\affiliation{%
  \institution{Center for Complex Networks and Systems Research; Luddy School of Informatics, Computing, and Engineering, Indiana University}
  \city{Bloomington}
  \state{IN}
  \country{USA}}
\renewcommand{\shortauthors}{xxx}
\newcommand{\pv}{\texttt{persona2vec}}

\begin{abstract}
Graph embedding techniques, which learn low-dimensional representations of a graph, are achieving state-of-the-art performance in many graph mining tasks.
Most existing embedding algorithms assign a single vector to each node, implicitly assuming that a single representation is enough to capture all characteristics of the node.
However, across many domains, it is common to observe pervasively overlapping community structure, where most nodes belong to multiple communities, playing different roles depending on the contexts.
Here, we propose \pv{}, a graph embedding framework that efficiently learns multiple representations of nodes based on their structural contexts.
Using link prediction-based evaluation, we show that our framework is significantly faster than the existing state-of-the-art model while achieving better performance.
\end{abstract}



\maketitle

\section{Introduction}

Graph embedding maps the nodes in a graph to continuous and dense vectors that capture relations among the nodes~\cite{Perozzi2014deepwalk,grover2016node2vec,Tang2015line}.
Resulting node representations allow direct applications of algebraic operations and common algorithms, facilitating graph mining tasks such as node classification~\cite{sen2008collective,Perozzi2014deepwalk}, community detection~\cite{fortunato2010community,yang2016modularity}, link prediction~\cite{grover2016node2vec} and visualization~\cite{Tang2015line}.
Most methods map each node to a single vector, implicitly assuming that a single representation is sufficient to capture the full characteristics of a node.

However, nodes often play multiple roles.
For instance, people have multiple roles, or ``personas'', across contexts (e.g. professor, employee, and so on) \cite{ahn2010link,coscia2014uncovering,leskovec2009community,leskovec2010empirical}.
Similarly, proteins and other biological elements play multiple functionalities \cite{palla2005uncovering,gavin2006proteome,ahn2010link}.
Another example is the polysemy of words when their relations are modeled with graphs; many words possess multiple meanings differentiated by the contexts \cite{chen2014unified,li2015multi,iacobacci2015sensembed}.
Explicit modeling of such multiplicity and overlapping clusters has been fruitful not only for community detection \cite{rosvall2014memory,coscia2014uncovering,epasto2017ego}, but also for improving the quality of embedding \cite{li2015multi,epasto2019single}.
Yet, with the scarcity of embedding methods embracing this idea, the full potential of this approach has not been properly explored.

In this paper, we propose \pv{}, a scalable framework that builds on the idea of ego-splitting \cite{epasto2017ego}, the process of identifying local structural contexts of a node via performing local community detection on the node's ego-network.
For each detected local community (role), we transform each node into multiple personas if there are multiple local communities to which the node belongs.
After the split, the original node is replaced by the new persona nodes that inherit the connection from each local community, producing a new persona graph.
Instead of separating a node's persona nodes from each other completely \cite{epasto2019single}, we add directed, weighted edges between personas to capture their origin.
In doing so, we allow the direct application of the existing graph embedding methods. In addition, we take an approach of considering persona-based learning as fine-tuning of the base graph embedding, achieving both efficiency and balance between information from the original graph and the persona graph.  Compared with the previous approach \cite{epasto2019single}, our framework is conceptually simpler to understand and practically easier to implement.
Furthermore, it achieves better performance in the link prediction tasks while being much faster.

In sum, we would like to highlight that our approach (1) drastically lowers the threshold for combining existing algorithms with persona splitting, (2) significantly improves the efficiency of the ego-splitting approach, while (3) consistently excelling the previous state-of-the-art.
Our implementation of \pv{} is publicly available at 
\url{https://github.com/jisungyoon/persona2vec}.

\section{Proposed Method: \pv{}}

\pv{} creates a \emph{persona graph}, where some nodes are split into multiple personas.
We then apply a graph embedding algorithm to the persona graph to learn the embeddings of the personas (see Fig.~\ref{fig:ego_split}).
Let us explain the method formally. Let $G = (V, E)$ be a graph with a set of nodes $V$ and a set of edges $E$. 
$|V|$ and $|E|$ denote the number of nodes and edges respectively.
Let $f\colon v \rightarrow \mathbb{R}^d$ be the embedding function that maps a node $v$ to a $d$-dimensional vector space ($d \ll |V|$).

\subsection{Refined Ego-splitting}

\begin{figure*}
    \centering
    \includegraphics[width=0.8\textwidth]{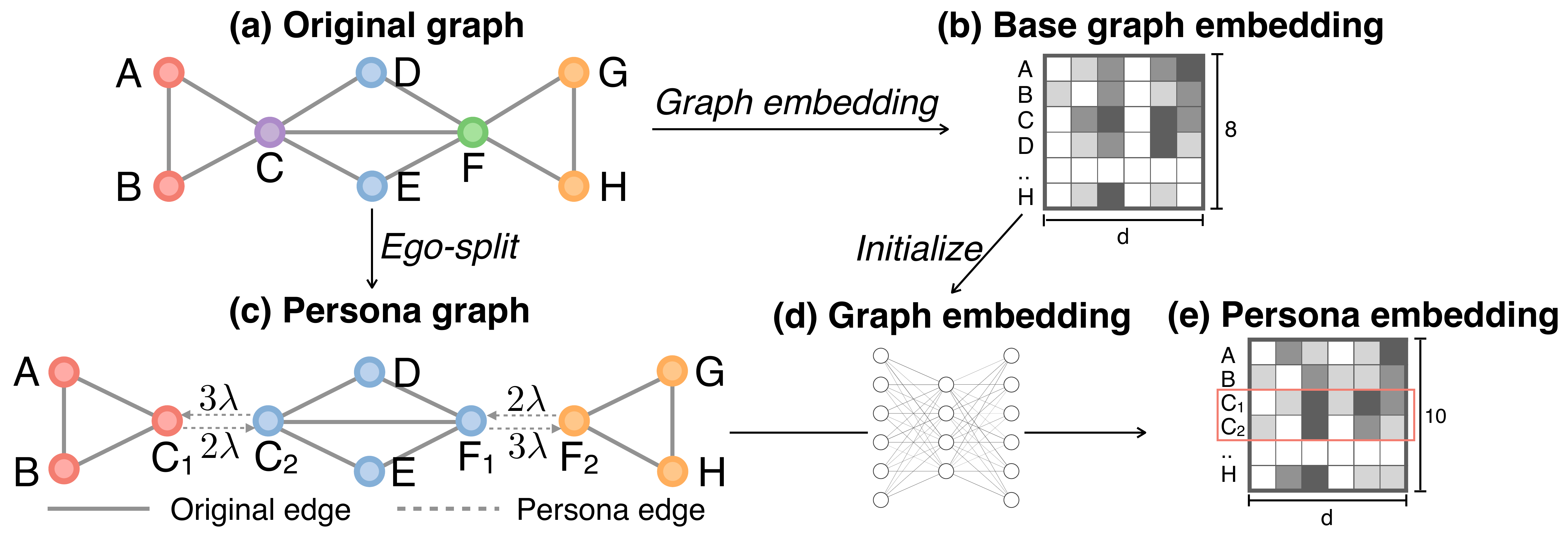}
    \caption{
    \textbf{Illustration of \pv{} framework.}
    (a) A graph with an overlapping community structure.
    (b) Graph embedding of the original graph is obtained first to initialize the persona embeddings.
    (c) Transform the original graph into a persona graph.
    Every edge in the original graph is preserved in the persona graph, while new directed persona edges with weight $\lambda k^{\text{o}}_i$ are added between the persona nodes.
    (d) Graph embedding is applied to the persona graph.
    (e) The final persona embedding where each persona node has its own vector representation.
    }
    \label{fig:ego_split}
\end{figure*}

We adopt and refine the ego-splitting method \cite{epasto2017ego,epasto2019single}.
For each node in the original graph, we first extract its ego graph, remove the ego, and identify the local clusters.
Every cluster in the ego graph leads to a new persona node in the persona graph (see Fig.~\ref{fig:ego_split}a, c).
For example, if we consider each connected component as a local community with a connected component algorithm, node $C$ in the original graph belongs to two non-overlapping clusters $\{A,B\}$ and $\{D,E,F\}$ in its ego-network.
Given these two clusters, in the persona graph, $C$ is split into $C_1$ and $C_2$ to represent the two roles in respective clusters.
$C_1$ and $C_2$ inherit the connections of $C$ from both clusters separately (see Fig.~\ref{fig:ego_split}c).
On the other hand, node $A$ only belongs to one ego cluster $\{B,C\}$, so it does not split into multiple personas.

\begin{algorithm}[t!]
	\caption{Refined ego-splitting for generating the persona graph. Case of the undirected graph} 
	\label{alg:ego_splitting} 
	\begin{algorithmic}[1]
		\Require{}
		\Statex{$G(V,E)$, the original graph}
		\Statex{$\lambda$, the weight factor for persona edges}
		\Statex{$\mathcal{C}$}, non-overlapping local clustering algorithm
		\Ensure{}
		\Statex{$G_P(V_P, E_P)$, the persona graph}
		\Statex{$V2P$, node to personas mapping}
		\Statex{$P2C$, persona to local cluster mapping}
		\algrule
		\Function{RefEgoSplit}{$G$}
		\For{{\textbf{each} $v_o \in V$}}
		    \State $P_{v_o} \leftarrow \mathcal{C}(v_o)$ \Comment{find local clusters of $v_o$}
		    \For{{\textbf{each} $p \in P_{v_o}$}}
		        \State Create $v_p$ \Comment{create persona nodes for local clusters}
		        \State Add $v_p$ to $G_P$
		        \State Add $v_p$ to $V2P(v_o)$
		        \State $P2C(v_p) \leftarrow p$
		    \EndFor
		\EndFor
		\For{{\textbf{each edge} $(v_i,v_j)$ \textbf{in} $E$}}
		    \State $w \leftarrow $ weight of edge
		    \For{{\textbf{each persona node} $v_p$ \textbf{in} $V2P(v_i)$}}
		        \For{{\textbf{each persona node} $v'_p$ \textbf{in} $V2P(v_j)$}}
		        \If{$v_i \in P2C(v'_p)$ and $v_j \in P2C(v_p)$}
    		        \State Add original edge $(v_p, v'_p, w)$ to $E_P$
    		        \State Add original edge $(v'_p, v_p, w)$ to $E_P$
    		    \EndIf
		        \EndFor
		    \EndFor
		\EndFor
		\State $k^o \leftarrow \text{out-degree sequence after adding original edges}$
		\For{{\textbf{each} $v_o \in V$}}
		    \For{{\textbf{each pair} $(v_i,v_j)$ \textbf{in} $V2P(v_o)$}}
                \State Add persona edge $(v_i, v_j, k^o_i \times \lambda)$ to $E_P$
                \State Add persona edge $(v_j, v_i, k^o_j \times \lambda)$ to $E_P$
        	\EndFor
		\EndFor

		\State{\Return $G_P(V_P, E_P), V2P$}
		\EndFunction
	\end{algorithmic}
\end{algorithm}

Any graph clustering algorithm can be employed for splitting a node into personas. The simplest algorithm is considering each connected component in the ego-network (sans the ego) as a cluster. This approach is fast and works well on sparse graphs. However, in dense graphs, ego-networks are more likely to form fewer connected component, thus other algorithms such as the Louvain method \cite{blondel2008fast}, Infomap \cite{rosvall2008maps}, and label propagation \cite{raghavan2007near} would be more appropriate. 

In previous studies, the personas get disconnected without retaining the information about their origin, creating isolated components in the splitting process \cite{epasto2017ego,epasto2019single}.
Because of this disconnectedness, common embedding methods could not be directly applied to the ego-split graph. A previous study attempted to address this issue by imposing a regularization term in the cost function to penalize separation of persona nodes originating from the same node \cite{epasto2019single}.

Here, instead of adopting the regularization strategy, we add weighted \emph{persona edges} between the personas, maintaining the connectedness between them after the splitting (see Fig.~\ref{fig:ego_split}c).
Because the persona graph stays connected, classical graph algorithms and graph embedding methods can now be readily applied without any modification.
As we will show later, our strategy achieves both better scalability and better performance.

In persona graph, we set the weights of the unweighted original edges as $1$ and tune the strength of the connections among personas with $\lambda$. Persona edges are directed and weighted, with weight $\lambda k^{\text{o}}_i$, where $k^{\text{o}}_i$ is the out-degree of the persona node after splitting (see Fig.~\ref{fig:ego_split}c). Assigning weight proportional to $k^{\text{o}}_i$ helps the random walker explores both the local neighbors and other parts of the graph connected to the other personas regardless of its out-degree $k^{\text{o}}_i$. 

Imagine node $u$, which is split into $n_p$ personas. Consider one of the personas $i$ with out-degree $k^{\text{o}}_i$ and persona edges with weight $w_i$. Then the probability $p_i$ that an unbiased random walker at $i$ visits neighbors connected with the original edge at the next step is $ \frac{k^{\text{o}}_i}{k^{\text{o}}_i + n_pw_i}$. If we set constant weight $w_i=\lambda$, then $p_i = \frac{k^{\text{o}}_i}{k^{\text{o}}_i + n_p\lambda} = \frac{1}{1 + \frac{n_p}{k^{\text{o}}_i}\lambda}$, which depends on $k^{\text{o}}_i$. A random-walker would not explore its local neighborhood if $n_p \gg k^{\text{o}}_i$, while the opposite happens when  $n_p \ll k^{\text{o}}_i$. Instead, assigning the weight proportional to $k^{\text{o}}_i$, namely $w_i=\lambda k^{\text{o}}_i$, removes such bias because $p_i=\frac{k^{\text{o}}_i}{k^{\text{o}}_i + n_p\lambda k^{\text{o}}_i} = \frac{1}{1 + n_p\lambda}$, which is independent of $k^{\text{o}}_i$.  Our experiments also show that using the out-degree yields better performance than assigning the identical weight to each persona edge. Our algorithm for refined ego-splitting is described in Algorithm~\ref{alg:ego_splitting}. Note that it can be generalized to the directed graphs.

\begin{algorithm}[t!]
	\caption{\pv{}. Our method for generating persona node embeddings.}
	\label{alg:persona2vec} 
	\begin{algorithmic}[1]
		\Require{}
		\Statex{$G(V,E)$, Original graph}
		\Statex{$d$, embedding dimension}
		\Statex{$\mathcal{\gamma}_b$, number of walks per node for base embedding}
		\Statex{$t_b$, random walk length for base embedding}
		\Statex{$w_b$, window size for base embedding}
		\Statex{$\mathcal{\gamma}_p$, number of walks per node for persona embedding}
		\Statex{$t_p$, random walk length for persona embedding}
		\Statex{$w_p$, window size for persona embedding}
		\Statex{$\alpha$, learning rate}		
		\Statex{\textsc{RefEgoSplit}, refined ego-splitting method}
		\Statex{$V2P$, node to personas mapping}
		\Statex{\textsc{EmbeddingFunc}, a graph embedding method e.g. DeepWalk, node2vec}
		\Ensure{}
		\Statex{$\Phi_{G_P}$, a $N_P \times d$ matrix with $d$-dimensional vector representations for all $N_P$ persona nodes}
		\algrule
		\Function{\pv{}}{$G$, $\textsc{EmbeddingFunc}$}
		\State{$G_{P}, V2P \leftarrow \textsc{RefEgoSplit}(G)$}
	    	
	    \State{$\Phi_{G} \leftarrow \textsc{EmbeddingFunc}(G,d,w_{b},\gamma_{b},t_{b},\alpha)$}
		\For{\textbf{each} $v_{o} \in V$}
		    \For{{\textbf{each persona node} $v_p$ \textbf{in} $V2P(v_{o})$}}
				\State $\Phi_{G_P}(v_p) = \Phi_{G}(v_o) $		
			\EndFor						
		\EndFor			
		
		\State{$\Phi_{G_P} \leftarrow \textsc{EmbeddingFunc}(G_{p},d,w_{p},\gamma_{p},t_{p},\alpha, \Phi_{G_P})$}
		
		\State{\Return $\Phi_{G_P}$}
		\EndFunction
	\end{algorithmic}
\end{algorithm}

\subsection{Persona graph embedding}

As explained above, any graph embedding algorithm that recognizes edge direction and weight can be readily applied to the persona graph.
Although we use \texttt{node2vec} as the embedding method here, other embedding methods can also be employed. We initialize the persona vectors with the vectors from the original graph before ego-splitting (see  Fig.~\ref{fig:ego_split}b) to leverage the information from the original graph structure.
Persona nodes that belong to the same node in the original graph are thus initialized with the same vector.
We then execute the embedding algorithm for a small number of epochs to fine-tune the embedding vectors with the information from the persona graph (see Fig.~\ref{fig:ego_split}).
Experiments show that usually only one epoch of training is enough.

Also, training the embedding on the persona graphs from scratch fails to yield comparable results. We find that initializing the embedding with the original graphs, i.e., our present method, consistently improves the performance, suggesting that mixing the structural information from both the original graph and the persona graph is crucial.
Our full algorithm is described in Algorithm~\ref{alg:persona2vec}.

\subsection{Complexity}

The persona graph is usually larger than the original graph, but not too large.
Node $u$ with degree $k_u$ may be split into at most $k_u$ personas.
In the worst case, the number of nodes in the persona graph can reach $O(|E|)$.
But, in practice, only a subset of nodes split into personas, and the number of personas rarely reaches the upper bound.
If we look at the persona edges, for a node $u$ with degree $k_u$, at most $O(k_u^2)$ new persona edges may be added.
Thus, the whole persona graph has at most $O(|V| \times k_{\text{max}}^2)$ or $O(|V|^3$) ($\because k_{\text{max}} \le |V|$) extra persona edges. If graph's degree distribution follows a power-law distribution $P(k) \sim k^{-\gamma} $, then $k_{\text{max}} \sim |V|^{1 / \gamma - 1}$. Hence, it could be $O(|V|^{\gamma + 1/\gamma - 1})$ and it is between $O(|V|^2)$ and $O(|V|^3)$ ($\because 2 \le \gamma \le 3$ in general). However, real graph tends to be sparse and $k_i\ll|V|$. If we further assume $k_i < \sqrt{|E|}$ holds for every node, then $\sum^{|V|}_{n=1} k_n^2 \leq \sum^{|V|}_{n=1} k_n \sqrt{|E|} = 2|E|\sqrt{|E|}$.
Under this assumption, the upper bound becomes $O(|E|^{3/2})$.
Similarly, with the scale-free condition, the upper bound could be  $O(|E||V|^{1 / \gamma - 1})$, which is between $O(|E||V|^{1/2})$ and $O(|E||V|)$.
Again, in practice, the number of persona edges is much smaller than this upper bound.
To illustrate, we list the number of nodes and persona edges in the persona graph for the graphs we use in this paper in Table~\ref{tab:statistics}.
All considered, the extra nodes and edges do not bring too much space complexity burden in practice.

Assessing the time complexity requires consideration of the two steps: ego-splitting and embedding.
The ego-splitting algorithm has complexity of $O(|E|^{3/2} + \sqrt{|E|} T(|E|))$ in the worst case, where $|E|$ is the number of edges in the original graph and $T(|E|)$ is the complexity of detecting the ego clusters in the graph with $|E|$ edges \cite{epasto2017ego}.
The embedding on the persona graph, which dominates the whole embedding procedure, has complexity $O(|V_p|\gamma t w d(1 + \log(|V_p|)))$ which is time complexity of \texttt{Node2vec}, where $|V_p|$ is the number of nodes, $\gamma$ is the number of random walkers, $d$ is the embedding dimension, and $w$ is the window size \cite{chen2018harp}.

\begin{figure}
    \centering
    \includegraphics[width=0.9\columnwidth]{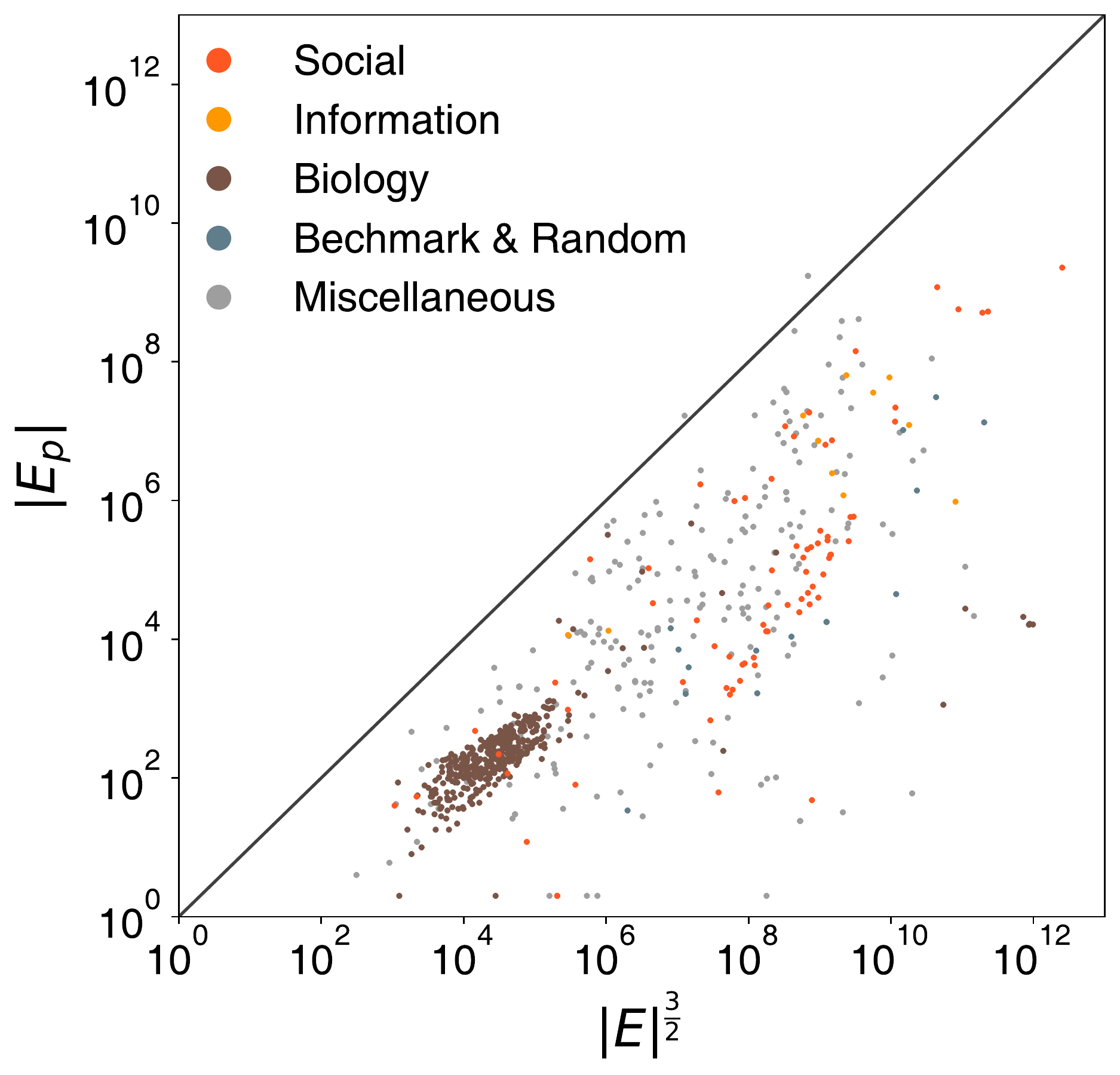}
    \caption{\textbf{Comparison of the the number of persona edges $|E_p|$ to the practical upper bound $|E|^{3/2}$.}}
    \label{fig:complexity}
\end{figure}

\begin{table*}
    \centering
    \caption{
    \textbf{Descriptive statistics in the graphs used in the evaluation.}
    We report the number of nodes $|V|$, number of edges $|E|$, number of nodes in the persona graph $|V_p|$, the ratio of $|V_p|$ over $|V|$, number of persona edges $|E_p|$ added in ego-splitting, and the ratio of $|E_p|$ over $|E^{3/2}|$ which is the upper bound of space complexity.
    } 
    \begin{tabular}{llrrrrrrr}
    \toprule
        Dataset & Type & $|V|$ & $|E|$ & |$V_p$| &|$V_p|/|V|$ & |$E_p$| &  $|E_p|/|E^{3/2}|$\\ 
    \midrule
        PPI & Undirected & 3,863 & 38,705 & 16,734& 4.34&132,932 &0.0175 \\  
        ca-HepTh & Undirected & 9,877 & 25,998& 16,071& 1.86 & 33,524 & 0.0800 \\ 
        ca-AstroPh & Undirected & 17,903 & 197,301& 25,706& 1.44 & 29,012 & 0.0003 \\ 
        wiki-vote & Directed & 7,066 & 103,633& 21,476& 3.04 &  118,020  & 0.0035 \\ 
        soc-epinions & Directed & 75,877 & 508,836& 220,332& 2.90 & 3,550,594  & 0.0098\\ 
    \bottomrule
    \end{tabular}
    \label{tab:statistics}
\end{table*}

The final complexity is $O(|E|^{3/2} + \sqrt{|E|} T(|E|)) + O(|V|\gamma t w d(1 + \log(|V|)))$. Removing the constant factors and assuming close-to-linear local community detection algorithm, the whole process has time complexity of $O(|E|^{3/2})$ with space complexity of $O(|E|^{3/2}$) if $k_i < \sqrt{|E|}$ holds. Complexity can be increased depending on the clustering algorithms on the ego-network.

To test the validity of our assumptions, we sample 1,000 graphs from a public network repository \cite{nr}.
We apply the refined ego-splitting with connected component algorithms on these samples and report the actual number of persona edges $|E_p|$ with respect to the practical upper bound $|E|^{3/2}$ in Fig. \ref{fig:complexity}, which shown that the actual number of persona edges $|E_p|$ rarely exceeds the tighter upper bound that we proposed and is usually orders of the magnitude smaller.

\subsection{Optimization}

Any kind of graph embedding method can be considered, for simplicity, we choose the classical random-walker based embedding method (e.g. \texttt{Node2Vec}, \texttt{DeepWalk}). In the model \cite{Perozzi2014deepwalk}, the probability of a node $v_i$ co-occurring with a node $v_j$ is estimated by
\begin{equation}
\label{eq:word2vec}
p(v_{i} \vert v_{j}) = \frac{\exp(\bm{\Phi'}_{v_i} \cdot \bm{\Phi}_{v_{j}})}{\sum_{k=1}^V \exp(\bm{\Phi'}_{v_k} \cdot \bm{\Phi}_{v_{j}})},
\end{equation}
where $\bm{\Phi}_{v_i}$ and $\bm{\Phi'}_{v_i}$ are the ``input'' and ``output'' embedding of node $i$. We use input embedding $\bm{\Phi}$ which is known to be more useful and more widely used. Denominator of eq.\ref{eq:word2vec} is computationally expensive \cite{yang2016modularity, cao2016deep} and there are two common approximations:  hierarchical softmax \cite{morin2005hierarchical} and negative sampling \cite{mikolov2013distributed}. We adopt negative sampling not only because it is simpler and popular but also because it show better performance as we see later.

\section{Case Study}

\begin{figure}
    \centering
    \includegraphics[width=0.9\columnwidth]{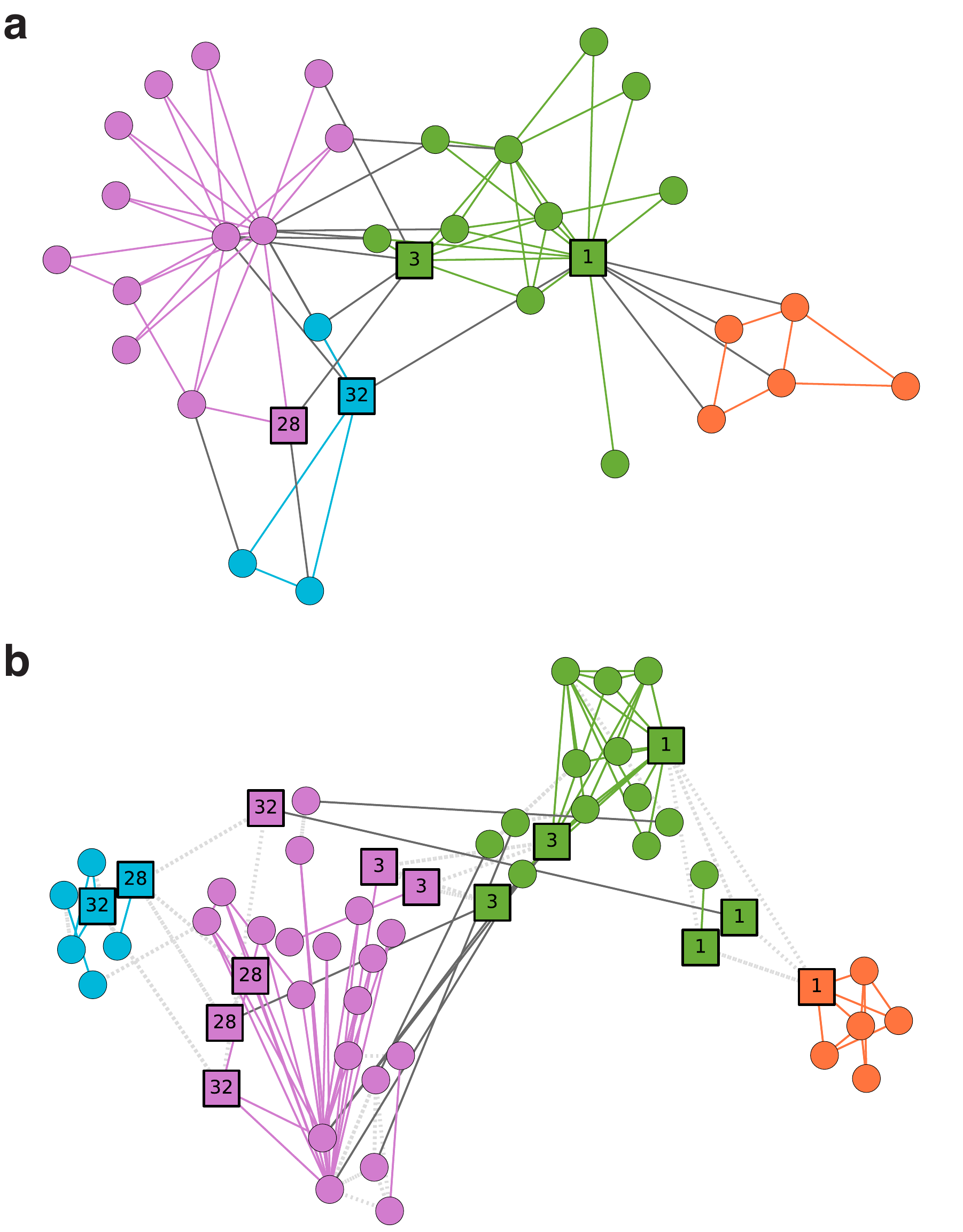}
    \caption{
        \textbf{Case Study: Zachary's Karate club network}
        (a) The Zachary's Karate club network with the force-atlas layout  \cite{zachary1977information}.
        Nodes are colored by communities detected by the Louvain modularity method \cite{blondel2008fast}.
        (b) The persona graph.
        Nodes are colored by k-means clusters \cite{macqueen1967some} from the embedding vectors.
        Coordinates of the persona nodes come from the 2-D projection of the embedding with t-SNE \cite{van2008visualizing}.
        Light grey lines represent the persona edges.}
    \label{fig:karate_example}
\end{figure}

Before diving into systematic evaluations, we provide two illustrative examples: Zachary's Karate club network and a word association network.

\paragraph{Case Study: Zachary's Karate club network}
We use the Zachary's Karate club network \cite{zachary1977information}, a well-known example for the community detection.
Nodes represent members of the Karate club, and edges represent ties among the members (see Fig. \ref{fig:karate_example}a).
Although it is often considered to have two large disjoint communities, smaller overlapping communities can also be seen, highlighted by nodes such as \texttt{1}, \texttt{3}, \texttt{28}, and \texttt{32}.
In Fig.~\ref{fig:karate_example}b, we present the persona graph of the network. \pv{} successfully recognizes these bridge nodes and place their personas in reasonable places.
Take node \texttt{1} for example. It splits into four persona nodes, which then end up in two different communities.
The orange and green communities are clearly separated as a result.

\begin{figure}
    \centering
    \includegraphics[width=1\columnwidth]{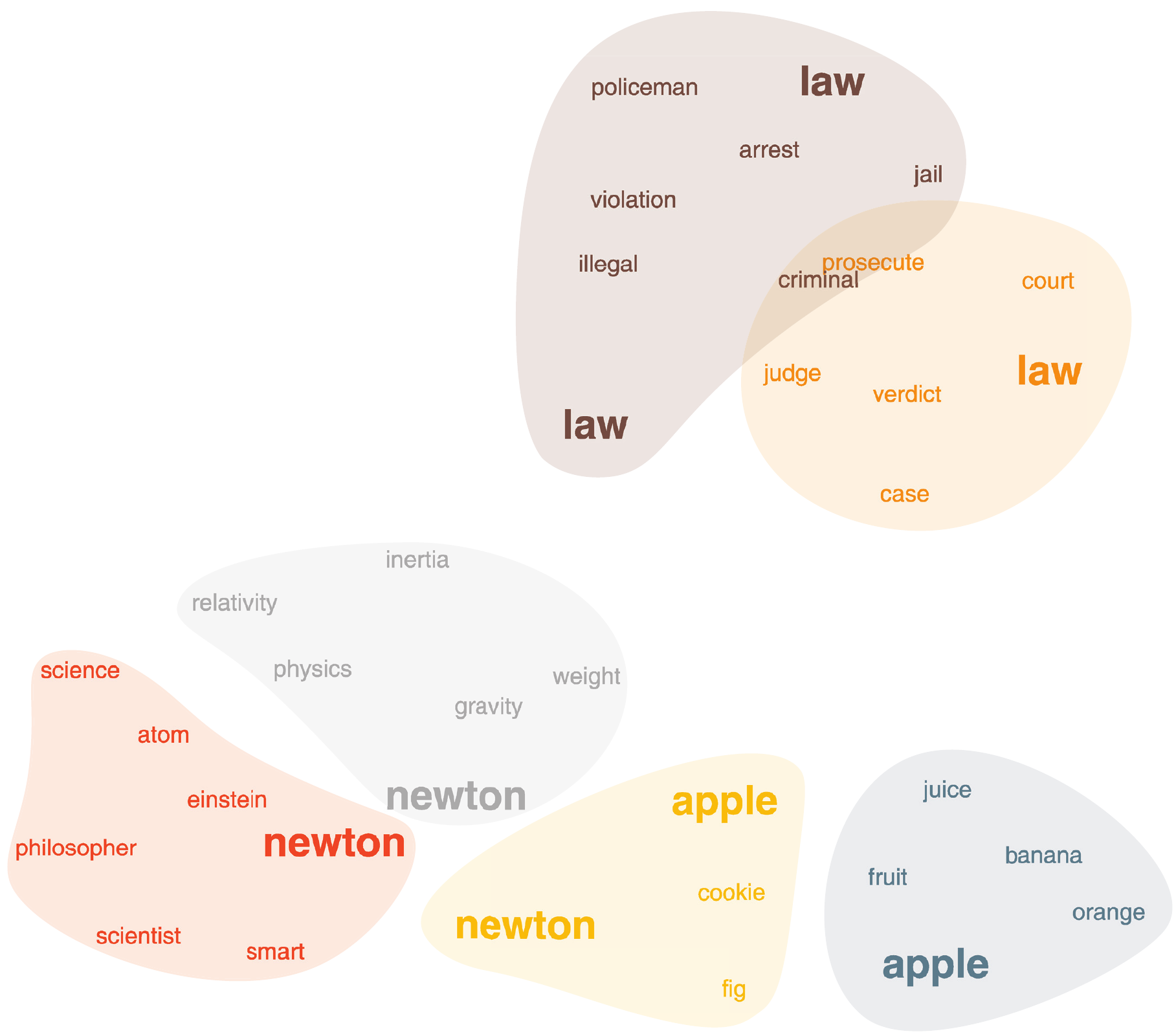}
    \caption{
        \textbf{The word association network, clusters around the word ``Newton''.}
        Coordinates of the words come from the 2-D projection of the embedding vectors with UMAP \cite{mcinnes2018umap}.
        Word colors correspond to the clusters obtained by k-means clustering \cite{macqueen1967some} on the embedding vectors.}
    \label{fig:word_example}
\end{figure}

\paragraph{Case Study: word association network}
Word association network captures how people associate words together (free association task). 
The dataset was originally assembled from nearly 750,000 responses from over 6,000 peoples.
Participants were shown 5,019 words and asked to write down the first word that sprang in mind and all the word pairs were collected with their frequency as the weights.
This dataset forms a weighted, directed graph of words that captures their multiple senses.
Although it is, in principle, possible to run our method on the original graph, for simplicity, we convert it into an undirected, unweighted graph by neglecting weight and direction \cite{ahn2010link}.
In Fig. \ref{fig:word_example}, we shows the \pv{} clusters around the word ``Newton''.
We use the Louvain method \cite{blondel2008fast} to split the personas of each word.
\pv{} successfully captures multiple contexts of the word ``Newton''.
For instance, the red persona is associated with ``scientists'' and ``philosopher'', grey one is linked to the physics, and yellow one is associated with ``apple'' (note that there is a cookie called ``(Fig) Newton'' in the U.S.).
Furthermore, \pv{} also captures different nuances of the word ``law'' that are related to the crime (brown cluster) and the legal concepts (orange cluster).

\section{Experiment Design}
\subsection{Link Prediction Task}

To systematically evaluate the performance and scalability of the \pv{} framework, we perform a link prediction task using real-world graphs \cite{grover2016node2vec,abu2017learning}.
Link prediction aims to predict missing edges in a graph with partial information, which is useful for many tasks such as suggesting new friends on social networks or recommending products.
It has been employed as a primary task to evaluate the performance of unsupervised graph embedding methods \cite{abu2017learning,zhang2018arbitrary}.

We follow the task setup from the literature \cite{grover2016node2vec,abu2017learning}.
First, the edge set of an input graph is divided equally and randomly into $E_{\text{train}}$ and $E_{\text{test}}$.
We then refine $E_{\text{test}}$ using a rejection sampling based on the criterion that, even when we remove all edges in $E_{\text{test}}$, the graph should be connected as a single component.
$E_{\text{train}}$ is used to train the models, and $E_{\text{test}}$ is used as positive examples for the prediction task.
Second, a negative edge set $E_{(-)}$ of non-existent random edges with the same size of $E_{\text{test}}$ are generated as negative examples for testing.
The performance of a model is measured by its ability to correctly distinguish $E_{\text{test}}$ and $E_{(-)}$ after being trained on $E_{\text{train}}$.
We then report ROC-AUC.

\begin{figure*}
    \centering
    \includegraphics[width=\textwidth]{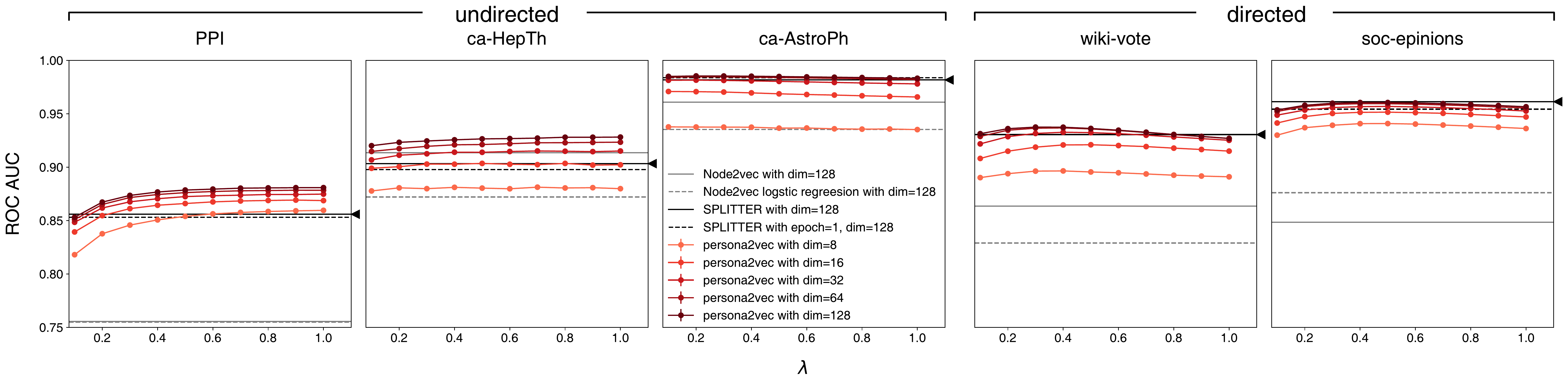}
    \caption{
    \textbf{Performance of \pv{} in the link prediction task.}
    The confidence intervals are all within the range of the markers.
    Given the same number of dimensions, \pv{} is always on par or better than \texttt{SPLITTER}
    }
    \label{fig:performance_result}
\end{figure*}

\begin{table*}
    \centering
    \caption{
        \textbf{Performance of \pv{} with $\lambda=0.5$.} All methods use $d=128$.  \texttt{Node2vec}* refers \texttt{Node2vec} with the logistic regression classifier, \texttt{SPLITTER}* refers \texttt{SPLITTER} with one epoch, and \pv{}* refers \pv{} with $\lambda=0.5$, our suggested default. Performance gain is performance difference between \texttt{Node2vec} and \pv{}*. We omit the standard error which is smaller than $10^{-3}$.
    } 
    \begin{tabular}{lrrrrr}
    \toprule
       Method & PPI & ca-HepTh & ca-AstroPh & wiki-vote & soc-epinions\\ 
    \midrule
        \texttt{Node2vec} & 0.756 & 0.914  & 0.961 & 0.864  & 0.849 $\pm$ 0.003\\
        \texttt{Node2vec}* & 0.755 $\pm$ 0.001 & 0.872   & 0.935  & 0.829 $\pm$ 0.001 & 0.871 $\pm$ 0.001\\
        \texttt{SPLITTER} & 0.856 & 0.903  & 0.982  & 0.931  & \textbf{0.961 $\pm$ 0.001} \\
        \texttt{SPLITTER}* & 0.853 & 0.898  & 0.984  & 0.931  & 0.954 $\pm$ 0.001 \\
        \textbf{\pv{}*} & \textbf{0.879} & \textbf{0.927} & \textbf{0.985} & \textbf{0.936} &\textbf{0.961} \\
    \midrule
        \textbf{\% Performance gain over \texttt{Node2Vec}} & 16\% & 1.4\% & 2\% & 8\% & 13 $\pm$ 0.3\%\\
    \bottomrule
    \end{tabular}
    \label{tab:performance_summary}
\end{table*}

\subsection{Datsets}

To facilitate the comparison with the state-of-the-art baseline, we use five graph datasets that are publicly available and previously used \cite{epasto2019single}.
We summarize them as follows.\\

\textbf{Undirected Graphs}
\begin{enumerate}[noitemsep,topsep=2pt]
    \item \textbf{PPI:} A protein-protein interaction graph of \emph{Homo sapiens} \cite{stark2006ppiorigin}.
    Nodes represent proteins and edges represent physical interactions between the proteins.
    \item \textbf{ca-HepTh:} A scientific collaboration graph.
    It represents the co-authorship among researchers from the Theoretical High Energy Physics field, derived from papers on arXiv.
    \item \textbf{ca-AstropPh:} A scientific collaboration graph. 
    It is similar to \textbf{ca-HepTh}, but from Astrophysics.
\end{enumerate}

\textbf{Directed Graphs}
\begin{enumerate}[noitemsep,topsep=2pt]
  \item \textbf{wiki-vote:}
  Each node is a Wikipedia user and a directed edge from node $i$ to node $j$ represents that user $i$ voted for user $j$ to become an administrator.
  \item \textbf{soc-epinions:} A voting graph from a general consumer review site \texttt{Epinions.com}. 
  Each node is a member and a directed edge from node $i$ to node $j$ means that member $i$ trusted member $j$.
\end{enumerate}

For \textit{PPI}, we use the prepossessed version from the \texttt{node2vec} project web page \cite{grover2016node2vec}, while other graphs are downloaded from the SNAP library homepage \cite{snapnets}.
We use the largest component of the undirected graphs and the largest weakly connected component of the directed ones. 
The statistics of all the graphs are reported in Table~\ref{tab:statistics}.

\subsection{Methods}

The state-of-the-art method in link prediction task is \texttt{SPLITTER} \cite{epasto2019single}, which also models multiple roles.
As reported in the paper, it outperforms various exiting reasonable algorithms ranging across non-embedding methods like \texttt{Jaccard Coefficient}, \texttt{Common Neighbors,} and \texttt{Adamic-Adar} as well as embedding methods like Laplacian \texttt{EigenMaps} \cite{belkin2002laplacian}, \texttt{node2vec} \cite{grover2016node2vec}, \texttt{DNGR} \cite{cao2016deep}, \texttt{Asymmetric} \cite{abu2017learning} and \texttt{M-NMF} \cite{wang2017community}.

Given the state-of-the-art performance of \texttt{SPLITTER}, for simplicity, we compare our framework with \texttt{SPLITTER} using the identical task setup and datasets.
In addition, because our method can be considered as an augmentation of a single-role embedding method, and because we use \texttt{Node2vec} as the base embedding method, we also employ \texttt{Node2vec}.
We run the link prediction task using the original authors' implementation of  \texttt{Node2vec} and \texttt{SPLITTER}.
The parameters are also kept consistent with the original paper.

\pv{} and \texttt{SPLITTER} have multiple representations on each node, which leads to non-unique similarity estimations between two nodes.
Hence, we define the similarity score of a pair of nodes on \pv{} as the maximum dot-product of embedding vectors between any pair of their personas.
We found that, among experiment with three aggregation functions \textit{min}, \textit{max}, \textit{mean}, the highest performance is achieved with \textit{max}, same with \texttt{SPLITTER} \cite{epasto2019single}.
For \texttt{SPLITTER}, we use maximum cosine similarity, following the author's note in their implementation.

\begin{figure*}
    \centering
    \includegraphics[width=0.9\textwidth]{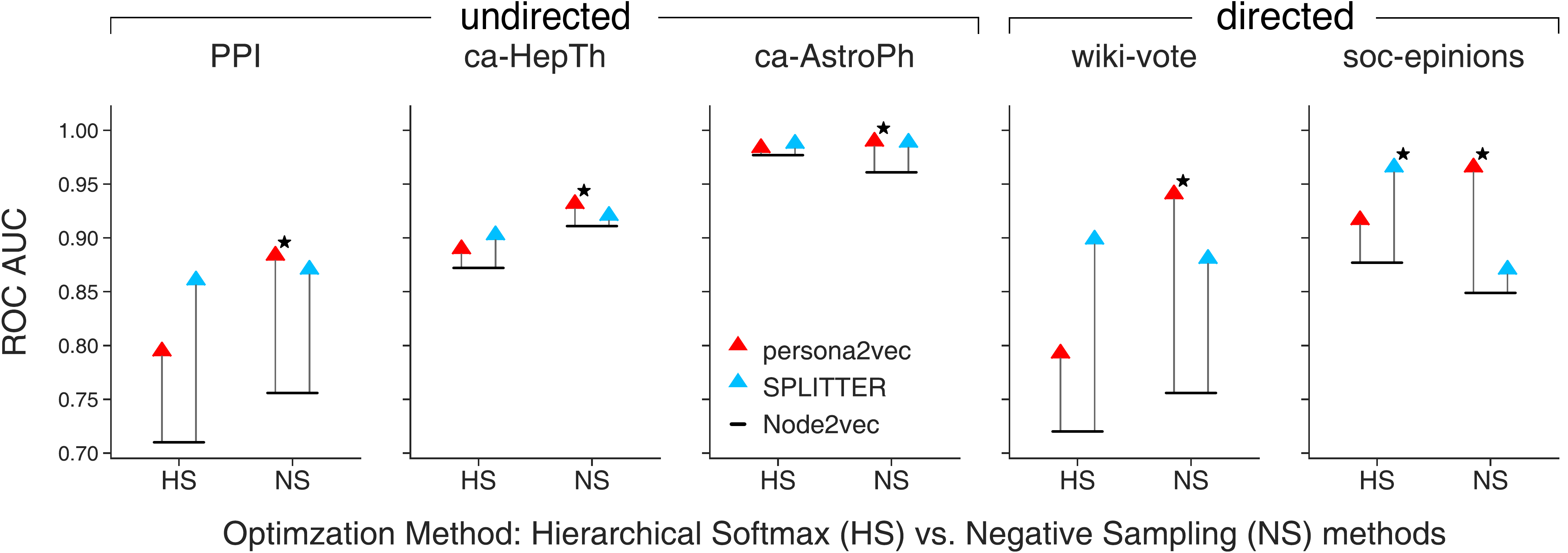}
    \caption{
    Comparison of link prediction performance between \pv{} and \texttt{SPLITTER} with different approximations. HS refers to the hierarchical softmax and NS refers to the negative sampling. The star marker indicates the best link prediction performance.
    }
    \label{fig:compare}
\end{figure*}

\begin{figure}
    \centering
    \includegraphics[width=0.95\columnwidth]{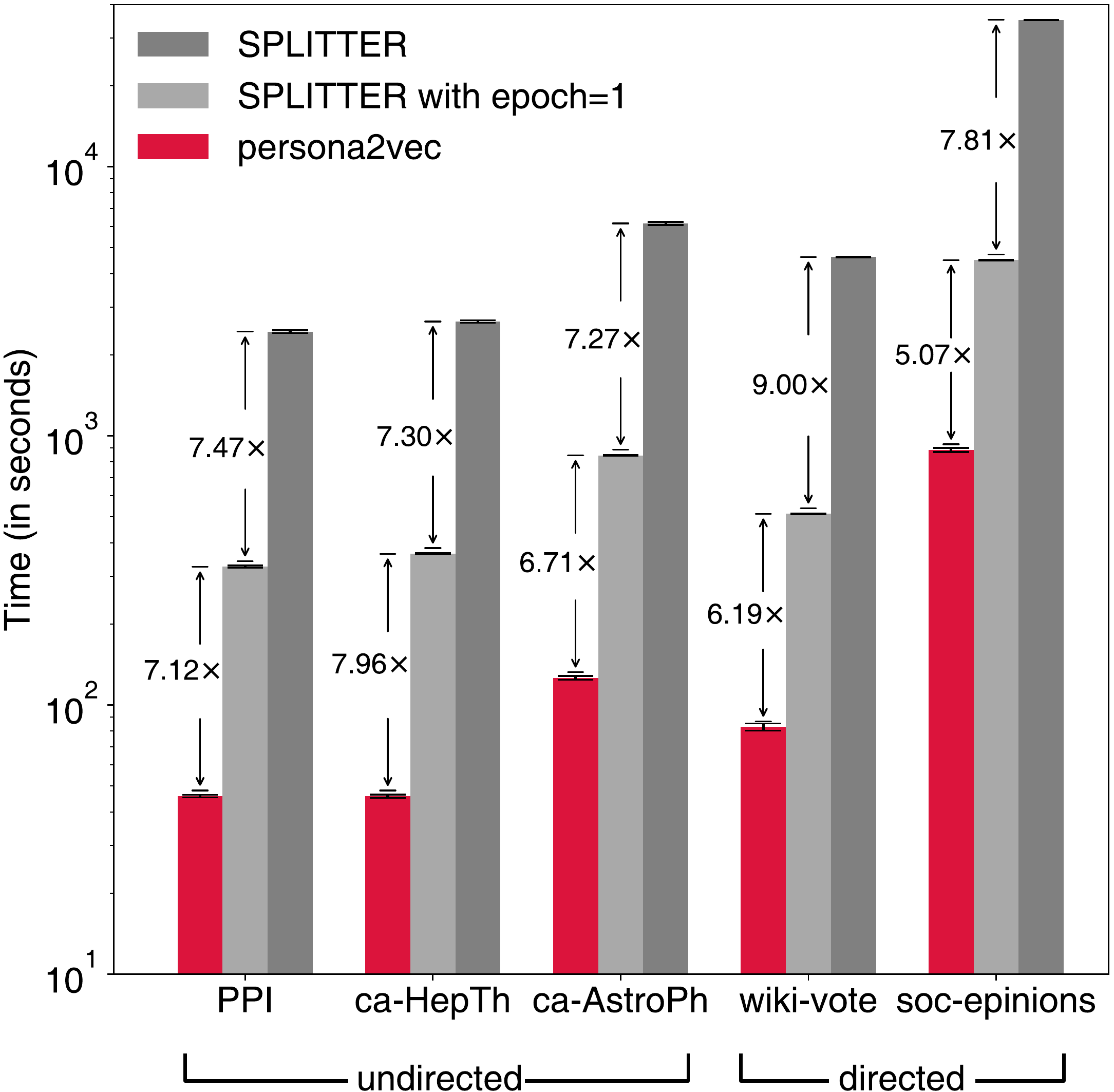}
    \caption{
    Comparison of elapsed time between \pv{} and \texttt{SPLITTER}.
    Speed gains by \pv{} are shown.
    }
    \label{fig:speed_result}
\end{figure}

\paragraph{\texttt{Node2vec} (baseline method)}

For \texttt{Node2vec}, we set random walk length $t=40$, the number of walks per node $\gamma=10$, random walk parameters $p = q = 1$, the window size $w=5$, and the initial learning rate $\alpha=0.025$. In the original paper, they learn an additional logistic regression classifier over the Hadamard product of the embedding of two nodes for the link prediction. In general, the logistic regression classifier improves the performance. Here, we report results on \texttt{Node2vec} with both dot products and the logistic regression classifier.

\paragraph{\texttt{SPLITTER} (baseline method)}

For \texttt{SPLITTER}, we use the same parameters in their paper \cite{epasto2019single} and \texttt{Node2vec} baseline. We use \texttt{node2vec} with random walk parameters $p = q = 1$.

\paragraph{\pv{} (our proposed method)}

We set the hyper-parameters of the original graph embedding with $t_{b}=40$, $\gamma_{b}=10$, $w_{b}=5$ (same as the baselines).
For the persona embedding, we set $t_{p}=80$, $\gamma_{p}=5$, $w_{p}=2$ to better capture the micro-structure of the persona graph.
The size of the total trajectories is determined by random walk length $t_{*}$ times number of walks per node $\gamma_{*}$, so we keep $t_{*} \gamma_{*}$  constant to roughly preserve the amount of information used in the embedding.
For both embedding stages, we use the $\alpha=0.025$, and \texttt{node2vec} with the random walk parameters $(p = q = 1)$ as the graph embedding function.

\subsection{Experiment Results}

Fig.~\ref{fig:performance_result} shows the link prediction performance of \pv{} in comparison with the baselines.
Overall, \pv{} yields superior performance across graphs and across a range of hyperparameter choice. We show that augmenting \texttt{Node2vec} by considering personas significantly improves the link prediction performance, evinced by the significant performance gain (see Table \ref{tab:performance_summary}).

As expected, larger dimensions lead to better performance, although \pv{} achieves reasonable results even with tiny embedding dimensions like 8 or 16.
We also show how the performance of \pv{} varies with $\lambda$.
For undirected graphs, larger $\lambda$ is beneficial but the trend saturates quickly.
For directed graphs, however, optimal performance is achieved with smaller values of $\lambda$.
In practice, we suggest starting with $\lambda=0.5$ as a default parameter because the overall variation brought by $\lambda$ is not substantial and even when the performance increases with $\lambda$, near-optimal performance can be achieved at $\lambda = 0.5$.

When compared with the \texttt{SPLITTER} baseline, \pv{} shows on par or better performances given the same embedding dimensions across a wide range of $\lambda$. We also report the performance summary for \pv{} with $\lambda=0.5$ (our suggested default) compared with the best baselines in Table \ref{tab:performance_summary}, which show that \pv{} outperforms the baseline consistently. Also, we report the ``performance gains'' from \texttt{Node2vec}, because we used \texttt{Node2vec} as the base embedding method and \texttt{persona2vec} can be considered an augmentation or fine-tuning of the base \texttt{Node2vec} vectors with local structural information. As shown, the persona-based fine-tuning significantly improved the performance. 

Also, we show the performance of both methods across different approximations: hierarchical softmax and negative sampling in Fig. \ref{fig:compare}.  We also found that cosine similarity consistently yields a better result with hierarchical softmax and dot product yields a better result with negative sampling across all methods. So, we use cosine similarity for hierarchical softmax results and use dot product for negative sampling results. We checked that both methods work well across the optimization method. We found that \pv{} tends to perform better with negative sampling and \texttt{SPLITTER} with hierarchical softmax. Nevertheless, \pv{} yields the best performance consistently.

In addition to the performance of the link prediction task, we also report the execution time of \pv{} and \texttt{SPLITTER} to compare their scalabilities in practice (see Fig. \ref{fig:speed_result}). Note that the reported execution time is on the link-prediction task, with half of the edges removed from the original graph.
\texttt{SPLITTER} runs the embedding procedures for 10 epochs by default in the original implementation, whereas \pv{} only runs for one epoch.
For a fair comparison, we also report the results of \texttt{SPLITTER} with one epoch of training. 
When being limited to only one epoch, \texttt{SPLITTER}'s performance slightly suffers on three graphs while it goes up or stays stable for the other two.

Nevertheless, \pv{} is more efficient---39 to 58 times faster than \texttt{SPLITTER} with $10$ epochs and five to eight times faster than \texttt{SPLITTER} with one epoch, while consistently outperforming both.
The most likely reason behind the drastic difference is the overhead from the extra regularization term in the cost function of \texttt{SPLITTER}, which \pv{} does not need. In sum, \pv{} outperforms the previous state-of-the-art method both in terms of scalability and link prediction performance.

\section{Related Work}

In addition to graph embedding, our work is closely related to the research of identifying overlapping communities in graphs.
Various non-embedding methods such as link clustering \cite{ahn2010link,PhysRevE.80.016105}, clique percolation \cite{palla2005uncovering}, and mixed membership stochastic blockmodel \cite{airoldi2008mixed} have been proposed.
Another thread of works focuses on using local graph structure to extract community information \cite{coscia2014uncovering,epasto2015ego,epasto2017ego}.
Specifically, Epasto et al. introduce the persona graph method for detecting overlapping communities in graphs \cite{epasto2017ego}, leveraging ego-network partition.
The combination of ego-network analysis and graph embedding methods is still rare.
An example is \texttt{SPLITTER} \cite{epasto2019single}, which we use as the baseline in this paper.
Instead of constraining the relations between personas with a regularization term, we propose a simpler and more efficient way of adding persona edges to the graph.

Our work is also related to the word disambiguation problem in word embedding.
Recently, word embedding techniques \cite{mikolov2013efficient,mikolov2013distributed,pennington2014glove} have been extensively applied to various NLP tasks as the vectorized word representations can effectively capture syntactic and semantic information.
Although some words have multiple senses depending on the context, the original word embedding methods only assign one vector to each word.
Li \textit{et al}. shows that embedding that is aware of multiple word senses and provides vectors for each specific sense does improve the performance for some NLP tasks \cite{li2015multi}.
For this issue, some utilize the local context information and clustering for identifying word sense \cite{reisinger2010multi,wu2015sense,neelakantan2015efficient}, some resort to external lexical database for disambiguation \cite{rothe2015autoextend,iacobacci2015sensembed,camacho2016nasari,chen2014unified,jauhar2015ontologically,pelevina2017making}, while some combine topic modeling methods with embedding \cite{liu2015topical,liu2015learning,cheng2015contextual,zhang2016improving}.
We adopt the idea of assigning multiple vectors to each node in the graph to represent different roles as well as exploiting local graph structure for the purpose.

\section{Conclusions}

We present \pv{}, a framework for learning multiple node representations considering the node's local structural contexts.
\pv{} first performs ego-splitting, where nodes with multiple non-overlapping local communities in their ego-networks are replaced with corresponding persona nodes.
The persona nodes inherit the edges from the original graph and remain connected by newly added persona edges, forming the persona graph.
Initialized by the embedding of the original graph, the embedding algorithm applied to the persona graph yields the final representations.
Instead of assigning only one vector to every node with multiple roles, \pv{} learns vectors for each of the personas.
With extensive link prediction evaluations, we demonstrate that \pv{} achieves the state-of-the-art performance while being able to scale better.
Moreover, our method is easy to comprehend and implement without losing any flexibility for incorporating other embedding algorithms, presenting great potential for applications. The possible combination with various algorithms provides vast space for further exploration.

As we know, the graph (relational) structure is ubiquitous across many complex systems, including physical, social, economic, biological, neural, and information systems, and thus fundamental graph algorithms have far-reaching impacts across many areas of sciences. Graph embedding, in particular, removes the barrier of translating methods to the special graph data structure, opening up a powerful way to transfer existing algorithms to the graphs and relational data. Furthermore, given that it is natural to assume overlapping clusters and their heterogeneous functionality in most real networks, multi-role embedding methods may find numerous applications in physical, biological, and social sciences.


\bibliographystyle{ACM-Reference-Format}
\bibliography{ref}






\end{document}